\documentstyle[equations,12pt]{article}
\newcommand{\ar}{\alpha}
\newcommand{\aar}{\bar{a}}
\newcommand{\bb}{\beta}
\newcommand{\gm}{\gamma}
\newcommand{\Gm}{\Gamma}

\newcommand{\dd}{\delta}
\newcommand{\ld}{\lambda}

\newcommand{\be}{\begin{equation}}
\newcommand{\ee}{\end{equation}}
\newcommand{\bea}{\begin{eqnarray}}
\newcommand{\eea}{\end{eqnarray}}
\newcommand{\bse}{\begin{subequations}}
\newcommand{\ese}{\end{subequations}}
\newcommand{\nn}{\nonumber}

\newcommand{\Dx}{\,_{q\!}D_x}
\newcommand{\Tx}{\,_{q\!}T_x}
\newcommand{\Tt}{\,_{q\!}T_t}
\newcommand{\Dt}{\,_{q\!}D_t}

\newcommand{\DDx}{\,_{q^2\!}D_{x^2}}

\newcommand{\DDa}{\,_{q^{2}\!}D_a}
\newcommand{\DDb}{\,_{q^{4}\!}D_b}
\newcommand{\TTa}{\,_{q^{2}\!}T_a}
\newcommand{\oR}{\bar{R}}
\begin{document}
%
%
\begin{center}
{\Large {\bf On a $q$-Deformation of the Discrete Painlev\'e I equation
and $q$-orthogonal Polynomials}}\\
\vspace{.6in}

{\large F.W. Nijhoff}$^\dagger$
\vspace{.1in}

{\small {\em Institute of Theoretical Physics, University of Amsterdam\\
Valckenierstraat 65, 1018 XE Amsterdam, The Netherlands}}\\
\vspace{.7in}

{\small October 1993}\\
\end{center}
\vspace{1in}

\centerline{ {\bf Abstract} }
\vspace{.5in}

I present a $q$-analog of the discrete Painlev\'e I equation,
and a special realization of it in terms of $q$-orthogonal polynomials.
\vspace{1in}

\footnotetext{$^\dagger\  $ \begin{tabbing}
Address after nov. 1: \= FB17-Mathematik,
Universit\"at-Gesamthochschule Paderborn \\
\> D-4790 Paderborn, Germany \end{tabbing} }

\pagebreak

\section{Introduction}

Recently, there is growing interest in difference versions of
Painlev\'e equations \cite{GR}. A first example,
the discrete Painlev\'e I equation (dPI),
which is a nonlinear non-autonomous ordinary difference equation of the
form
\be
\frac{ n + \ar }{ R_n} \ =\  \bb \left( R_{n+1} + R_n + R_{n-1}\right)
\, + \, \gm    ,
\label{eq:dPI} \ee
appeared in the theory of matrix models for
2-dimensional quantum gravity, cf. \cite{BK}-\cite{DS}. There it occurs
as lowest non-trivial term in the so-called `string equation'.

Other discrete Painlev\'e equations were subsequently found. A discrete
version of Painlev\'e II was obtained in \cite{PS}, starting from
unitary rather than hermitean one-matrix models, and independently
in \cite{NP} in connection with similarity reductions of integrable
lattice equations. Next, using the new method of singularity
confinement, \cite{GRP}, discrete analogues of Painlev\'e III,
IV and V, were found in \cite{RGH}. As the singularity confinement
approach is based on what is still a conjecture (namely, the
non-propagation of spontaneous singularities in integrable mappings),
proofs of integrability of these equations had still to be provided.
For dPI and dPII, the situation was clear from the beginning, because
their very construction (either by the approach of orthogonal
polynomials or by the similarity approach) led to the existence of
isomonodromic deformation problems for these equations, cf.
\cite{FIK} respectively \cite{NP}.
However, for the new discrete equations, dPIII-dPV, a priori no
Lax pair was known. For dPIV and dPV the situation is still open,
but in \cite{PNGR} a Lax pair for dPIII was presented (among new
isomonodromy problems for dPI and dPII, and their variants).
Interestingly enough, this Lax pair is a discrete isomonodromic
deformation problem of a linear {\em q-difference equation}, rather
than of a differential equation. It is to my knowledge the first time
that $q$-difference equations arise in connection with Painlev\'e
equations. To study the asymptotics of such type of systems, it is
necessary to go back to the historic works of the Birkhoff school on
this subject, cf. e.g. \cite{Carm}-\cite{LeCaine}, that unfortunately
seem to have fallen into oblivion since the 1940's. Although it is
relatively easy to find special solutions for the dPIII equation,
cf. \cite{GS},
the analysis of the corresponding $q$-difference system seems quite
complicated. 
As the theory of isomonodromic deformations for $q$-difference systems
still needs to be developed, it would, be useful to have at ones
disposal some slightly less complicated systems than the one for
dPIII, in order to study the features of this kind of analysis.

In this note, I would like to introduce for that purpose  a new
discrete Painlev\'e type of system, namely a $q$-deformed version of
the dPI equation (\ref{eq:dPI}). This is probably the simplest
$q$-deformed transcendent, although it is naturally written as a third
order difference equation rather than a second-order one like
(\ref{eq:dPI}). Nevertheless, the $q$-deformed discrete Painlev\'e I
equation, carries --like dPIII--  an
underlying monodromy problem of $q$-difference type. This system
is a natural candidate to be investigated from an analytical point of
view, namely by constructing asymptotic solutions and the connection
coefficients between them and formulating an inverse (Riemann-
Hilbert type) of problem along the lines laid out in \cite{Birk}.
The isomonodromic deformation can then be used to `solve' the
corresponding Painlev\'e transcendent. I will leave this program
for a future publication, and in this note I will confine myself to
establishing
a connection between the $q$-deformed Painlev\'e I and $q$-orthogonal
polynomials, very much along the lines in which this connection was
made in the one-matrix models. However, in order to do do this, we
will notice that one needs to  modify slightly the equation, namely
by introducing one term higher in the string equation. Furthermore,
by making this connection we will be naturally led to some
discrete-time analogue of the Volterra equation, the similarity
reduction of which is the $q$-deformed Painlev\'e I equation.

\section{$q$-Hermite Polynomials}

$q$-Analysis, \cite{Exton,GasRah} is very fashionable nowadays,
mainly because of
its use in quantum groups and the investigations
on $q$-special functions in connection with their representation
theory, cf. e.g. \cite{Koorn}-\cite{FR}.  In this note I would like to
draw attention to a new class of $q$-special functions, namely
$q$-difference deformed versions of Painlev\'e transcendents, in
particular a $q$-deformed version of the dPI equation (\ref{eq:dPI}).
It is well-known that the latter equation arises as the lowest
non-trivial term in the string equation
for one-matrix models via orthogonal polynomials. A special subcase,
namely if $\ar=\bb=0$ in (\ref{eq:dPI}), gives rise to the Hermite
polynomials.
In seeking
a $q$-deformed version it is natural, therefore, to start from the
basic $q$-Hermite polynomials, defined e.g. in \cite{Exton}. Let us
first review these, mainly for the purpose of fixing the notations.

First, let me recall the definition of the $q$-exponential
function\footnote{ In contrast with the
original exponential function, the $q$-exponential function $e_q^x$
has a finite radius of convergence $R=|1-q|^{-1}$ as $|q|<1$, and
simple poles at $x=(1-q)^{-1}q^{-j}\ $ for $\ j=0,1,2,\dots\ $. }
\be
e_q^{x}\,=\,\sum_{n=0}^{\infty} \frac{x^n}{(n)_q !}\      \ , \     \
\left( e_q^x\right)^{-1}\,=\,e_{q^{-1}}^{-x}\   ,
\label{eq:exp}
\ee
introducing also the $q$-numbers  $\ (n)_q\equiv \frac{q^n-1}{q-1}\ $.
Sometimes it is useful to work with Andrew's notation, \cite{GasRah},
\[ (a;q)_{\infty} \equiv \prod_{j=0}^{\infty} ( 1 - q^ja )\    \ ,
\    \ (a;q)_n \equiv \frac{ (a;q)_{\infty} }{ (aq^n;q)_{\infty} }
\  ,  \]
in terms of which we have e.g.
\be e_{q^{-1}}^x\,=\,( (q-1)x ; q)_\infty\     \ ,\   \ (\ q<1\ )\  .
\label{eq:eq} \ee
Now we can introduce the basic $q$-Hermite polynomials by means of
the generating function
\be
S_q(x,t)\,=\, e_q^{xt} e_{q^{-2}}^{\aar t^2}\,=\,
\sum_{n=0}^{\infty}\, \frac{t^n}{(n)_q !} H_n(x)\   ,
\label{eq:S}  \ee
Furthermore, introducing the $q$-differentiation and the $q$-dilatation
operator
$$\Dx\,f(x)\equiv \frac{f(qx)-f(x)}{(q-1)x}\     \ ,\    \
\Tx f(x)\equiv f(qx)\   , $$
and recalling that the $q$-exponential obeys the difference rules
$$\Dx\, e_q^{ax}\,=\,ae_q^{ax}\   \ ,\    \
\Dx\, e_{q^{-1}}^{ax}\,=\,ae_{q^{-1}}^{qax}\ ,   $$
one can derive for $S_q$ the following relations
\bse \label{eq:DS} \bea
\Dx S_q &=& t S_q\   , \label{eq:DSa}  \\
\Dt S_q &=& \left( x + (2)_q\aar t\right) \Tx^{-1}\Tt S_q\  ,
\label{eq:DSb} \eea \ese
leading to the relations
\bse \label{eq:DH} \bea
\Dx H_n &=& (n)_q H_{n-1}\   , \label{eq:DHa}  \\
H_{n+1} &=& xq^n\Tx^{-1} H_n\,+\,\aar (2)_q (n)_q q^{n-1} \Tx^{-1}
H_{n-1}\  . \label{eq:DHb} \eea
The last equation can be converted into
\be
H_{n+1}\,=\,x H_n\,+\,\aar (2)_{q^{-1}} (n)_q q^n H_{n-1} \  ,
\label{eq:DHc}
\ee
whereas, from (\ref{eq:DHa}) and (\ref{eq:DHb}) one can also derive the
second order $q$-difference equation
\be
q^{-n}(n)_q H_n\,=\,x \Dx\Tx^{-1} H_n\,+\,\aar (2)_{q^{-1}}
\Dx \Tx^{-1} \Dx H_n \  ,
\label{eq:DHd} \ee  \ese
i.e. the $q$-analogue of the Hermite equation.
Using the fact that $\Tx^{-1}\Dx = q\Dx \Tx^{-1}$, and that
$\Dx = (2)_qx\DDx$, we can convert the last equation into
\be
\Dx\left[ e_{q^{-2}}^{-aq^{-2}x^2} \Tx^{-1} \Dx H_n \right]\ +\
a(2)_{q^{-1}} q^{-n} (n)_q e_{q^{-2}}^{-ax^2} H_n\ =\ 0  \   ,
\label{eq:DDH} \ee
identifying $\ a = \left(-\aar (q^{-1}+1)^2\right)^{-1}\ $, which we
take to be real positive.  From
(\ref{eq:DDH}) it is clear that the basic $q$-Hermite polynomials
are orthogonal with respect to the Jackson integral
$$ \int_{-c}^{c}\,f(x) d_qx \equiv (1-q)c
\sum_{n=0}^{\infty}\,q^n \left[ f(q^nc) + f(-q^nc) \right]\   \ ,
\     \  c=\left( (1-q^{2})a\right)^{-1/2}\    \ ,\    \
(0<q<1) \  , $$
with weight function $w_q(x)=e_{q^{-2}}^{-ax^2}$.
Introducing the following representation for the $q$-Gamma function,
\cite{Exton,GasRah},
\be  \Gm_q(x)\, =\, q^{-x}\,\int_0^{1/(1-q)} t^{x-1}
e_{q^{-1}}^{-t} d_qt\, =\, \frac{ (q;q)_{\infty} }{ (q^x;q)_{\infty} }
(1-q)^{1-x}\   ,   \label{eq:Gm}  \ee
in terms of which we can express the normalization constant, we have
the orthogonality condition
\be
\int_{-c}^{c}\,d_qx\, e_{q^{-2}}^{-ax^2} H_n(x)
H_m(x)\ =\ 2\frac{ q^{\frac{1}{2} n(n+1)}
(n)_q!}{(2)_{q^{-1}}^{n+1} a^n\sqrt{a}} \Gm_{q^{2}}(\frac{1}{2})
\dd_{n,m} \   . \label{eq:orth}
\ee
Finally, the dependence on the variable $a$ (or, equivalently,
on $\aar$) is significant. In fact, from
\[ \DDa S_q\,=\,\frac{t^2}{(2)_q^2a^2} S_q\  , \]
we derive
\be
\DDa H_n\, =\, \frac{(n)_q (n-1)_q}{ (2)_q^2 a^2} H_{n-2}\   .
\ee

\section{A $q$-deformed Painlev\'e I Equation}

Comparing the Lax pair for dPI, cf. e.g. \cite{FIK}, with the one for
the $q$-Hermite polynomials, namely eqs. (\ref{eq:DHa}) and
(\ref{eq:DHc}), it is natural to pose the following $q$-linear system
\bse \label{eq:qLax} \bea
xP_n(x) &=& P_{n+1}(x)\,+\,R_n P_{n-1}(x)\  ,   \label{eq:qLa}   \\
\Dx P_n(x) &=& A_n P_{n-1}(x)\,+\,B_n P_{n-3}(x)\  ,   \label{eq:qLb}
\eea \ese
the compatibility of which leads to the set relations
\bse \bea
A_{n+1} &=& qA_n + 1 \  , \label{eq:a}  \\
B_{n+1} + R_n A_{n-1} &=& q\left( A_nR_{n-1} + B_n \right) \  ,
\label{eq:b}  \\
R_nB_{n-1} &=& qB_nR_{n-3} \  . \label{eq:c}
\eea \ese
Eqs. (\ref{eq:a}) and (\ref{eq:c}) are readily solved as
\be
A_n\,=\,(n)_q + \ar q^n\      \ ,\    \
B_n\,=\,\bb q^{-n} R_nR_{n-1}R_{n-2}\   .
\ee
Inserting these expressions into (\ref{eq:b}) we obtain the
following third-order nonlinear non-autonomous ordinary
difference equation
\be
\bb q^{-n} \left( q^{-1} R_{n+1}\,-\, q R_{n-2} \right)\,=\,
q\frac{ (n)_q + \ar q^n}{R_n}\,-\,\frac{ (n-1)_q +
\ar q^{n-1}}{R_{n-1}}\  ,  \label{eq:qPI}
\ee
which we will refer to as qPI.  The connection is made clear by
rewriting (\ref{eq:qPI}) in the form
\be
q^{-n}\left[ \gm\, + \,\bb q^{-2} \left( R_{n+1} + R_n + R_{n-1}\right)
\, + \, \bb (q^{-2} -1) \sum_{j=-\infty}^{n-2}\,R_j \right] \ =\
\frac{ (n)_q + \ar q^n }{ R_n} \   ,
\label{eq:qPIint} \ee
namely by performing one (formal) `integration'. The infinite sum term
in
(\ref{eq:qPIint}) will clearly disappear in the limit $q\rightarrow 1$,
in which case we immediately recover dPI , (\ref{eq:dPI}).

In \cite{Spiri} a $q$-deformed
Painlev\'e equation  related to a special subcase of the discrete
PII has been presented which shows some similarity to eq.
(\ref{eq:qPI}). It is not clear, however, whether their equation can be
reduced to (\ref{eq:qPI}) by coalescence. Furthermore, the approach
in \cite{Spiri} does not focuss on isomonodromic deformation problems,
but rather starts from factorizing Schr\"odinger operators.
What is interesting about
(\ref{eq:qPI}), apart from being probably the simplest $q$-deformed
transcendent that is now at our disposal, is that there is a
connection with $q$-orthogonal polynomials, along the same lines as
for the original dPI. However, in order to establish this connection
one has to modify slightly eq. (\ref{eq:qPI}) as we shall see.

\section{$q$-Orthogonal Polynomials}

We look for an explicit realization  of the qPI equation in
terms of $q$-orthogonal polynomials, in much the same way as
the original discrete Painlev\'e I equation (\ref{eq:dPI})
is solved by means of orthogonal polynomials. It is this
solution that gives rise to a connection with discrete hermitian
one-matrix models.

For this purpose let us generalize the orthogonality conditions
(\ref{eq:orth}) to
\be
\int\,d_qx\, e_{q^{-2}}^{-a x^2}
e_{q^{-4}}^{-b x^4} P_n(x) P_m(x)\ =\ h_n \dd_{n,m} \   .
\label{eq:north} \ee
For the integration limits in the Jackson integral (\ref{eq:north})
we take the smallest zero of the weight function
$\ w_q(x)=e_{q^{-2}}^{-ax^2}e_{q^{-4}}^{-bx^4}\ $, i.e. $\pm c$, where
$\ c = {\rm min}\,\left\{ \left( (1-q^{2})a\right)^{-1/2},
\left( (1-q^{4})b\right)^{-1/4}\right\}\ $, (for fixed positive $a,b$,
taking $0<q<1$) to ensure positivity of the measure.
The polynomials are supposed to
be normalized such that
\[ P_n(x)\,=\,x^n + \cdots\ \  \ \Rightarrow \  \
\Dx P_n\,=\,(n)_qP_{n-1} + \cdots \   .  \]
Of course one may go further and introduce in (\ref{eq:north})
a weight factor depending on an arbitrary
number of $q$-exponential factors, but I will refrain from doing so
in the present note.

{}From the orthogonality condition (\ref{eq:north}) we can derive the
isomonodromy problem (\ref{eq:qLa}), with $\ R_n=h_n/h_{n-1}\ $ (as
in the usual case). However, (\ref{eq:qLb}) needs to be modified
as follows from the following relation for the weight function
\[ \Dx w_q(x)\,=\, -\left[ (2)_qax\,+\,(4)_qbx^3\,+
\,(q-1)(2)_q(4)_qabx^5 \right] w_q(qx)\   , \]
which indicates that one has to push already to fifth order terms
in order to take into account second and fourth order coupling
constants $a,b$. This means that eq.
(\ref{eq:qLb}) is not applicable here, and performing the
calculation
\bea
\int &&d_qx\, w_q(x) \left[ (\Dx P_n) P_m
\,+\,P_n (\Dx P_m) \right]\,+\,
(q-1)\int\,d_qx\, w_q(x)x(\Dx P_n)(\Dx P_m)\nn \\
&&=\, \int\,d_qx\, w_q(qx) (\Tx P_n)(\Tx P_m)
\left[ (2)_qax\,+\,(4)_qbx^3\,+\,(q-1)(2)_q(4)_qabx^5 \right]\  ,
\label{eq:calc}
\eea
we are led to a linear problem consisting of (\ref{eq:qLa})
together with
\be
\Dx P_n(x)\,=\,A_n P_{n-1}(x)\,+\,B_n P_{n-3}(x)\,+\,C_n P_{n-5}\  .
\label{eq:qLc}
\ee
Compatibility will now lead to
\bse  \label{eq:Leqs} \bea
A_{n+1} &=& qA_n + 1 \  , \label{eq:aa}  \\
B_{n+1} + R_n A_{n-1} &=& q\left( A_nR_{n-1} + B_n \right) \  ,
\label{eq:bb}  \\
C_{n+1} + R_n B_{n-1} &=& q\left( B_nR_{n-3} + C_n \right) \  ,
\label{eq:cc}  \\
R_nC_{n-1} &=& qC_nR_{n-5} \  . \label{eq:dd}
\eea \ese
Eqs. (\ref{eq:aa}) and (\ref{eq:dd}) again are readily solved as
\be  \label{eq:AC}
A_n\,=\,(n)_q + \ar q^n\      \ ,\    \
C_n\,=\,\gm q^{-n} R_nR_{n-1}R_{n-2}R_{n-3}R_{n-4}\   ,
\ee
and  from (\ref{eq:bb}) and (\ref{eq:cc}) we get the system
\bse \label{eq:qqPI} \bea
q\frac{ (n)_q + \ar q^n}{R_n}\,-\,\frac{ (n-1)_q +
\ar q^{n-1}}{R_{n-1}} &=&
\gm q^{-n} \left( q^{-1} R_{n+1}\bar{B}_{n+1}\,-\, q R_{n-2}
\bar{B}_n \right)\   , \\
\bar{B}_n\,-\,\bar{B}_{n-1} &=& q^{-2}R_{n+1}\,-\,R_{n-4}\   ,
\label{eq:qqb}
\eea \ese
in which $B_n=q^{-n}\gm R_nR_{n-1}R_{n-2} \bar{B}_n$. Of course,
one can derive from (\ref{eq:qqPI}) a closed sixth order
difference equation by eliminating the $\bar{B}_n$. This is not so
illuminating, and I will not give the formula.
However, by formally solving (\ref{eq:qqb}) as
\[ \bar{B}_n\,=\,\bb\gm^{-1}\,+\,q^{-2}\left(
R_{n+1}+R_n+R_{n-1}+R_{n-2}+R_{n-3}\right)\,+\,(q^{-2}-1)
\sum_{j\leq n-4}\,R_j\  , \]
we see that the case of (\ref{eq:qPIint}) is included for
$\gm\rightarrow 0$.

On the other hand, from the orthogonality condition, performing the
calculation (\ref{eq:calc}) one derives relations between
the coefficients $A_n$, $B_n$, $C_n$. For $A_n$ and $C_n$ one
finds again (\ref{eq:AC}) for the special choice $\ar=0$
respectively $\gm=(1-q^{-1})q^4(2)_{q^{-1}}(4)_{q^{-1}}ab$.  Furthermore,
for $B_n$ we obtain
\bea
\bar{B}_n &+& (q-1)q^{-2n+4}\gm R_{n-2}R_{n-3}
R_{n-4}\bar{B}_{n-2}   \nn \\
&=& q^2(4)_{q^{-1}}b\gm^{-1}\,+\,q^{-2}\left( R_{n+1} + R_n + R_{n-1} +
R_{n-2}\right) \,+\, q^{2-n} R_{n-3}\   ,  \label{eq:BB}
\eea
which turns out to be consistent with (\ref{eq:qqPI}), and in
addition we have the following equation
\bea
q^n \frac{(n)_q}{R_n} &=& (2)_{q^{-1}}a\,+\,
(4)_{q^{-1}}b\left( R_{n+1} + R_n \right)
\,-\,\gm q^{-n} R_{n-1}R_{n-3}   \nn \\
&+& \gm q^{-4} \left[ R_{n+1}\left( R_{n+2} + R_{n+1} + R_n
\right)\,+\, R_n\left( R_{n+1} + R_n + R_{n-1}\right) \right] \nn \\
&+& \gm \bar{B}_n \left[ \bar{B}_{n+2}\,-\,q^2 (4)_{q^{-1}}b\gm^{-1}
\,-\, q^{-2}\left( R_{n+3} + R_{n+2} + R_{n+1} + R_{n} \right) \right]
\nn \\
&+& (1-q)\gm^2 q^{-2n} R_{n}R_{n-1}R_{n-2}R_{n-3}R_{n-4}\  ,
\label{eq:RR}
\eea
which is the actual $q$-deformed string equation (at least for
the lower order terms), identifying $\bb=q^2(4)_{q^{-1}}b$.
Also eq. (\ref{eq:RR}) is consistent with
(\ref{eq:qqPI}), which can be checked explicitely by tedious
algebra. Thus, the orthogonality is consistent with the compatibility
condition, which in turn implies that the $q$-orthogonal
polynomials are indeed solutions of the system (\ref{eq:qLc}).

\section{Discrete-Time Flows}

The dependence on the `coupling constants' $a,b,\dots$ as time-flow
parameters in (\ref{eq:north}) is significant, and will lead to a
discrete-time version of the Volterra system. In fact
for the polynomials $P_n$, we can derive from (\ref{eq:north})
the $q$-difference time-evolutions of the form
\bse \label{eq:DPab} \bea
\DDa P_n &=&  Q_n P_{n-2}\  , \label{eq:DPaa}  \\
\DDb P_n &=&  U_n P_{n-2}\,+\, V_n P_{n-4}\  ,\label{eq:DPbb}
\eea \ese
where the coefficients $Q_n$, $U_n$ and $V_n$ are determined by the
orthogonality condition. If we would have included additional higher-order
exponents in the weight function
of (\ref{eq:north}) this would lead to a discrete-time Volterra
hierarchy. Let us illustrate this by working out explicitely the lowest
order time-flow, namely in terms of the variable $a$. From
(\ref{eq:DPaa}) together with (\ref{eq:qLa}) we obtain the
relations
\bse \label{eq:DRQ} \bea
\DDa R_n &=&  Q_n\,-\,Q_{n+1} \  , \label{eq:DR}  \\
(\TTa R_n)Q_{n-1} &=&  Q_n R_{n-2}\  ,\label{eq:DQ}
\eea \ese
from which one can derive that
\be \label{eq:Q}
Q_n\,=\,c \oR_n\oR_{n-1}\     \ , \      \
\oR_n = R_n +(1-q^{-2})aQ_{n+1}\   \ ,\    \ c={\rm constant}\   .
\ee
This then leads immediately
to the following exact discrete-time $q$-deformation of the
Volterra equation
\be
(\DDa \oR_n)\,=\,c\left[ \oR_n\oR_{n-1}\,-\,q^2\TTa(\oR_{n+1}\oR_n)
\right]\  .\label{eq:Volt}
\ee
It is slightly surprising
that the exact integrable discretization of the Volterra systems turns
out to take a form which resembles closely the original
continuous-time equation, provided one considers the proper variables
$\oR_n$. Furthermore, solving (\ref{eq:DQ}) by taking $\ Q_n=(\TTa h_n)/
h_{n-2}\ $, recalling that $\ R_n=h_n/h_{n-1}\ $, we are led to the
following equation for $h_n$
\be  h_{n-1}(\TTa h_n)\,=\,\left[ h_n\,+\,(1-q^2)a(\TTa h_{n+1})\right]
\left[ h_{n-1}\,+\,(1-q^2)a(\TTa h_n)\right] \  , \label{eq:h}
\ee
where taking $c=1$ is consistent with the orthogonality.
On the other hand, from the orthogonality condition (\ref{eq:north})
one can derive by a calculation similar to (\ref{eq:calc})
\be
-\DDa h_n\,=\,\TTa \left( h_{n+1} + \frac{h_n^2}{h_{n-1}}\right)\,
+\,(1-q^2)a \frac{(\TTa h_n)^2}{h_{n-2}}\   , \label{eq:hh}
\ee
which can be shown to be consistent with (\ref{eq:h}).
Let me mention that eq. (\ref{eq:h}) is a special subcase of a universal
lattice equation derived some years ago in \cite{NQC}.

Of course, the above derivation can be extended to the
higher-order time-variables, e.g. starting from (\ref{eq:DPbb}).
In general this will lead to more complicated systems and we will
abstain in this note from presenting their derivation.
What is important to note, however, is that for all the coefficients
$a,b,\dots$ in the weight function (which in certain contexts are
interpreted as coupling constants), the $q$-deformation in
terms of the spectral parameter $x$ leads necessarily also to a
$q$-deformation in terms op these higher-order `time'-variables
which in turn leads to integrable discrete-time systems like the
discrete Volterra system of (\ref{eq:Volt}).

\section{Discussion}

I have shown that it is fairly straightforward to obtain a $q$-analogue
of the discrete Painlev\'e I equation by straightforwardly
$q$-deforming the continuous isomonodromy problem for discrete PI.
In this
way one obtains eq. (\ref{eq:qPI}) which is associated with probably
the simplest isomonodromic deformation problem of a linear
$q$-difference equation. Having this system at our disposal,
we can now seriously embark on the more serious problem of
investigating the full asymptotics for the corresponding transcendents.

Furthermore, I have put forward a connection with $q$-orthogonal
polynomials. For this one needs to consider
a slightly more complicated system including 5th order terms in the
``string equation''. The connection with orthogonal polynomials is
interesting, because on the one hand it establishes what is, in fact,
a special similarity solution of the discrete-time Volterra
hierarchy. On the other hand, it suggests a discretization
of the Hermitean one-matrix model, which --on the continuum level--
is well known to be related to orthogonal polynomials of Painlev\'e
type. It is suggestive, therefore, to try to trace back
the $q$-orthogonal polynomials to the corresponding matrix models.
If one were naive, one would be
tempted to write down an expression of the form
\be Z_N(a,b)\,=\,\int_{N\times N} [d_qH] e_{q^{-2}}^{-a trH^2}
e_{q^{-4}}^{-b trH^4}\   , \label{eq:Z} \ee
for the $q$-deformation of the partition function of the
Hermitean one-matrix model, in which $[d_qH]$ denotes a proper
$q$-analogue of the Haar measure. However, eq. (\ref{eq:Z}) cannot be
correct, because the $q$-exponents do not decompose in a natural way
for linear combinations in their arguments, nor do the Jackson
integrals allow
for arbitrary changes of variables. What one needs is an interpretation
of an expression of the form of (\ref{eq:Z}) which should lead --after
integration over the ``angle variables'' associated with an Hermitean
matrix $H$-- to a multiple Jackson integral over the eigenvalues of
$H$ of the form
\be
Z_N\,=\,\int \left[ \prod_{i=1}^N\,d_q\ld_i\,e_{q^{-2}}^{-a\ld_i^2}
e_{q^{-4}}^{-b\ld_i^4}\right]\,\prod_{i<j=1}^N (\ld_i - \ld_j )^2\   .
\label{eq:ZZ}
\ee
It is this expression for a `partition function' that has a direct
connection with the qPI equation of section 4. However, how to arrive
at this expression starting from a (discrete) matrix
integral needs some further study which is beyond the scope of
this paper.

\section*{Acknowledgement}

The author is grateful for the hospitality of the Institute for
Theoretical Physics of the University of Amsterdam,
and acknowledges financial support from the Dutch Organization of
Scientific Research (NWO) during his visit. He is also
indebted to Hans Capel, Vyacheslav Spiridonow and Luc Vinet for some
stimulating discussions.

\end{document}